\documentclass[pral,twocolumn,showpacs,preprintnumbers,amsmath,amssymb]{revtex4}% Physical Review B
\usepackage{graphicx}% Include figure files
\usepackage{dcolumn}% Align table columns on decimal point
\usepackage{bm}% bold math

\begin{document}

\title{Complementarity and Afshar's experiment}

\author{Aur\'{e}lien Drezet }
\email{aurelien.drezet@uni-graz.at} \affiliation{Institut f\"ur
Experimentalphysik, Karl Franzens Universit\"at Graz,
Universit\"atsplatz 5 A-8010 Graz, Austria}

\date{\today}

\begin{abstract}
In this article we criticize the experiment realized by S.~Afshar
[Proc.~SPIE 5866, 229-244 (July 2005)]. We analyze Bohr's
complementarity and show that the interpretation proposed by
Afshar is misleading.
\end{abstract}

\pacs{03.~65.~Ta, 32.~80.~Lg, 07.~79.~Fc} \maketitle

In an article recently submitted for publication \cite{Afshar}
S.~Afshar claims to have successfully realized an experiment
disproving the complementarity principle of N.~Bohr
\cite{Bohr1,Heisenberg,Feynman,Wheeler}. Obviously this result, if
verified, would constitute a serious attack against the orthodox
interpretation of quantum mechanics (known as the Copenhagen
interpretation). However the aim of this comment is to criticize
briefly the result obtained by Afshar and to show that
the interpretation presented in \cite{Afshar} is completely misleading.\\
In the following we will not consider the exact set up used in [1]
but a slightly modified version. \begin{figure}[h]
\includegraphics[width=3.5in]{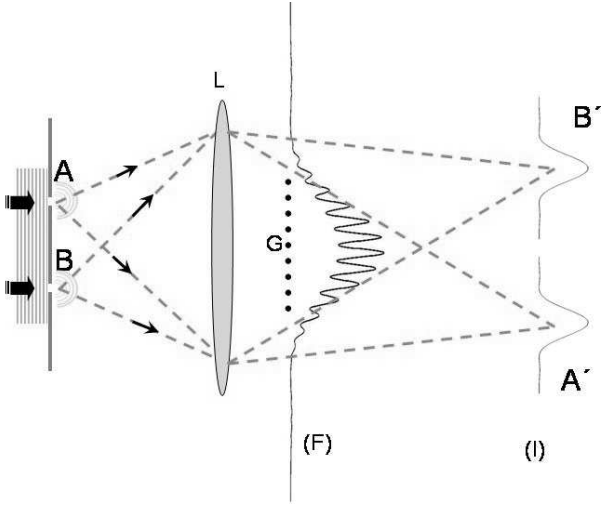}\\
\caption{A idealized version of Afshar's experiment. This
experiment is very close from a gedanken experiment proposed
originally by Wheeler\cite{Wheeler}. }\label{fig2}
 \end{figure}However since we are only
interested in the general principle this idealized version of [1]
, which is sketched in Fig.~1, is completely sufficient for our
purpose. As preliminary analysis, we consider the Young
double-pinhole experiment in the case of a single photon wave
train impinging normally on the apertures plane. The photon state
immediately behind the holes evolves into a sum of two diffracted
waves:
$\Psi\left(\mathbf{x}\right)=\Psi_{A}\left(\mathbf{x}\right)+\Psi_{B}\left(\mathbf{x}\right)$,
where $\Psi_{A,B}$ are single aperture wave functions reducing to
two narrow peaks located at $\mathbf{x}_{B,A}=\pm
d/2\mathbf{\hat{x}}$ in the apertures plane. The two spherical
waves don't need to have necessary the same amplitude and we note
$\Psi\left(\mathbf{x}\right)\sim A f\left(\mathbf{x}
-\mathbf{x}_{A}\right)+B f\left(\mathbf{x} -\mathbf{x}_{B}\right)$
the photon wave function in the apertures plane. We have typically
$f\left(\mathbf{x}\right)\sim\delta^{2}\left(\mathbf{x}\right)$
i.~e.~a two dimensional Dirac function. Neglecting the overlap
between the two waves in the holes the intensity
$I\left(\mathbf{x}\right)\propto|\Psi\left(\mathbf{x}\right)|^{2}$
equals approximately  $|A f\left(\mathbf{x}
-\mathbf{x}_{A}\right)|^{2}+|B f\left(\mathbf{x}
-\mathbf{x}_{B}\right)|^{2}$. If we put a detection screen just
behind the apertures plane we expect to see two narrow peaks of
relative intensities $P_{A}= |A|^{2}/\left( |A|^{2}+
|B|^{2}\right)$ and $P_{B}= |B|^{2}/\left( |A|^{2}+
|B|^{2}\right)$. $P_{A,B}$ define then probabilities to record a
photon in $A$ or $B$. In this context we can define the
distinguishability
\begin{equation}
D=|P_{A}-P_{B}|= ||A|^{2}-|B|^{2}|/\left( |A|^{2}+
|B|^{2}\right).\label{dist}
\end{equation}
The meaning of $D$ is clear if we consider the case $P_{A}=P_{B}$
corresponding to $D=0$. Indeed if we detect a photon in the far
field of the aperture planes (i.~e. in the plane where we expect
to see fringes) we can not say from which hole cames from the
photon: the two holes are indistinguishable. Oppositely if
$P_{B}=1=D$ we are sure that the photon detected in the far field
came from $B$ and not from $A$. In order to obtain a mathematical
formulation of wave-particle duality we must consider the
interference pattern recorded in the far-field. To do that we can
introduce a converging lens behind the apertures plane. The effect
of this lens on the photon can be mathematically described by
Fresnel's diffraction theory. We consider here a lens with a
gaussian transmission
$T\left(\mathbf{x}\right)=e^{-\left(\mathbf{x}/\sigma\right)^{2}/2}$.
We omit the irrelevant $z$ dependent factor in the wave function
(the $z$ axis being the optical axis) and we deduce
\begin{eqnarray}
\Psi\left(\mathbf{X}\right)\propto
Ae^{-k^{2}\alpha\left(\mathbf{X}/P'
+\mathbf{x}_{A}/P\right)^2}\nonumber\\+Be^{-k^{2}\alpha\left(\mathbf{X}/P'
+\mathbf{x}_{B}/P\right)^2}
\end{eqnarray} for
every point $\mathbf{X}=[X,Y],z$ located after the lens. We have
$\alpha^{-1}=2\left(1/\sigma^2-ik\epsilon\right)$ (where
$k=2\pi/\lambda$ is the photon wave vector) and
$\epsilon=1/P+1/P'-1/f$ which depends on the focal length $f$, the
distance $P$ between the lens and the pinholes, and the distance
$P'$ between the lens and the observation plane. In the focal
plane of the lens we obtain\begin{eqnarray}
\psi\left(\mathbf{X}\right)=N\cdot[Ae^{-ik P\left(\mathbf{X}/f
+\mathbf{x}_{A}/P\right)^2/2}\nonumber\\+Be^{-ikP\left(\mathbf{X}/f
+\mathbf{x}_{B}/P\right)^2/2}],
\end{eqnarray} with the approximation $1/\sigma^{2}\ll k/P$. The
intensity is thus $I\left(\mathbf{X}\right)\propto
1+V\cos{\left(kd\mathbf{\hat{x}}\cdot\mathbf{X}/f+ \phi\right)}$
where the visibility $V$ and the phase shift $\phi$ are defined by
\begin{eqnarray}
V=2|A||B|/\left(|A|^{2}+|B|^{2}\right)\nonumber\\
\phi=\arg{\left(A\right)}-\arg{\left(B\right)}.\end{eqnarray} Such
visibility is equivalently defined by the relation
$V=\left(I_{max}-I_{min}\right)/\left(I_{max}+I_{min}\right)$. It
should be noted that we could like Afshar observe fringes in front
of the lens where the intensity reads
$I\left(\mathbf{X}\right)\propto
1+V\cos{\left(kd\mathbf{\hat{x}}\cdot\mathbf{X}/p+ \phi\right)}$
and $V,\phi$ are the same than previously. If $p\gg f$ it is
indeed easier to observe the fringes in front of the lens.\\
It exists a strong relation between $V$ and $D$ since we
have\begin{eqnarray} D^{2}+V^{2}=1 &\textrm{(duality
relation\cite{Englert})}.\label{englert}
\end{eqnarray}
It means that the observation of a perfect interference pattern
with unit visibility $V=1$ implies a total indistinguishability
$D=0$. Oppositely if we can  predict from which hole the photon
came from (i.~e.~$D=1$) then we can't have interference:
i.~e.~$V=0$. This is already a formulation of the principle of
complementarity. In this example this principle can be enounced in
the following form: since we can not absorb a photon twice we must
select between either observing the photon in the interference
plane or in the aperture plane. If, with both apertures open and
$|A|=|B|$, we detect several photons in the interference plane,
then we will (by statistical accumulation) record an interference
pattern with unit visibility. However in counterpart we can not
build up, \emph{by using the same events}, the statistical
distribution of particles in the aperture plane because we have no
information for that. The only case in which we can build the two
patterns with the same particles is if we consider a single
aperture (i.~e.~$D=1$). But in that case the fringes visibility
obviously equals zero. We considered here a simple experiment but
the validity of this duality relation is in fact much more
general\cite{Englert}. Eq.~\ref{englert} is applicable in cases
involving quantum entanglement with a ``which- path'' detector.
The aim of this comment is however not to discuss the concept of
complementarity in its general sense and we refer to the original
literature for that purpose \cite{Bohr1,Englert,Scully}.\\
The previous canonical example was slightly modified by S.~Afshar.
In his experiment Afshar considered actually what happens if we
record the photon in the image plane of the lens (where
$\epsilon=0$ from geometrical optics). We have
\begin{eqnarray}
\Psi\left(\mathbf{X}\right)=\propto
Ae^{-k^{2}\sigma^{2}\left(\mathbf{X}/P'
+\mathbf{x}_{A}/P\right)^2/2}\nonumber\\+Be^{-k^{2}\sigma^{2}\left(\mathbf{X}/P'
+\mathbf{x}_{B}/P\right)^2/2}
\end{eqnarray}
which give us (in the limit $k\sigma d/P\ll 1$) two well separated
gaussian spots centered on the geometrical image points $A'$ and
$B'$ of the pinholes $A$ and $B$. The intensity is then
approximatively given by
\begin{eqnarray}
I\left(\mathbf{X}\right)\propto
|A|^{2}e^{-k^{2}\sigma^{2}\left(\mathbf{X}/P'
+\mathbf{x}_{A}/P\right)^2}
\nonumber\\+|B|^{2}e^{-k^{2}\sigma^{2}\left(\mathbf{X}/P'
+\mathbf{x}_{B}/P\right)^2} .\label{eq1}
\end{eqnarray}
At the limit of the infinite lens we obtain two dirac pics
$I\left(\mathbf{X}\right)\propto|A|^{2}\delta^{2}\left(\mathbf{X}/P'
+\mathbf{x}_{A}/P\right)+|B|^{2}\delta^{2}\left(\mathbf{X}/P'
+\mathbf{x}_{B}/P\right)$. Eq.~\ref{eq1} implies that the number
of particle detected at the center of the image $A'$ (respectively
$B'$) is directly proportional to the number of particles
potentially detectable in $A$ (respectively $B$). We can as
previously case introduce the same relative intensity $P_{A,B}$
and define the distinguishability $D$ in an identical way. Obviously the duality relation
Eq.~\ref{englert} is still valid.\\

At that point a comment is necessary. Indeed the usual formulation
of Bohr's principle says that any devices capable of determining
the path taken by a particle trough the double-aperture must
destroy the interference. The term path very often used in
discussions concerning the wave-particle duality is however
ambiguous and non univocally defined. More precisely it is not
obvious in the present experiment (with both pinholes open) that a
photon coming, say from $A$, should necessarily finish its journey
in $A'$. This axiom is strongly dependent of the ontological model
of hidden variable considered. In particular the ontological model
of Bohm when generalized to photons leads to difficulties (see for
example \cite{Drezet,Scully2}). For this reason we must limit
references to the word ``path'' in quantum mechanics. In fact we
can enounce Bohr's principle independently of any ontological
hypothesis. Considering the previous double-pinhole example we can
say that it is impossible by using the same events to build up the
statistical distribution associated with two complementary
observable of a same quantum state. This is clear since the wave
function in the interference plane is the fourier transform of the
wave function in the aperture plane. Recording photons in the
image plane of the lens change nothing to the analysis: if we
detect events in the image $A'$ and $B'$ we can build up the
statistical distribution of photons in the aperture plane. However
we can not build up the interference pattern. To do that we need
to know in which point would be detected each individual events in
the focal plane. However this exclude to use the same photons for
building up the image spots $A'$ and $B'$ since a photn can not be
absorbed twice. Once again this statement is independent of any
ontological models considered: it is valid if we think in term of
Bohm's trajectory, spontaneous collapse or
whatever you want if this agrees with quantum mechanics. \\
In the same context it is important to remark that Afshar's use of
Eq.~\ref{englert} is misleading and confusing. Indeed he remarked
that if we consider only one of the image we can define a
population of photons for which $D'=1$ and not $D=0$. This is
correct but then we can not use the relation $D'^{2}+V^{2}=1$
since we refer to two different ensembles of particles (only half
of the events are detected in $A'$ or $B'$). If we record photons
in the focal plane of the lens (i.~e.~ in the interference plane)
we can not attribute a label to photons saying in which of the two
images $A'$ or $B'$ it will or would finish its journey (if he was
not already recorded in the focal plane). Afshar didn't respected
these limitations and found that $D'^{2}+V^{2}=1+1=2$. This is
mathematically true but this is not the duality relation and can
not be considered as a formulation of Bohr's principle.

In the last steps of his experiment Afshar introduced absorbing
wires at the interference minima locations. If the wires are
sufficiently thin we don't expect an effect on the image spots
$A'$ and $B'$ when both pinholes are open. In his particular set
up Afshar observed actually a small reduction of the intensity of
$R\sim 0.1\%$ due to absorbtion by the wires. This intensity
reduction is much smaller than the reduction $R~\sim 6\%$ expected
for a uniform flow of energy in the interference plane (this can
be experimentally tested by opening only one of the two pinholes).
This is clearly an indication that interference occurs in the
focal plane of the lens. Afshar then claims that he recorded both
the statistical distribution of particles in the image plane and
the interference pattern.

A first and intuitive critic against this interpretation is that
the grid of wires acts like a grating disturbing the motion of the
photons. It is indeed visible with the grid in place and by
successively opening and closing one of the two pinhole, say $A$
(B being always open), that the shape of the spot $A'$ is modified
\cite{Afshar}. This obviously proves that a contribution coming
from $B$ is going in $A'$ when both holes are open. One could then
argue that the which path information is lost during this
interaction with the grating. A simple way to see this effect is
to consider a simple wire or diffracting object located in the
Fourier plane (i.~e.~ the focal plane) of the lens.  This object
induce a scattered field which can be in principle written $\simeq
\alpha\Psi(\mathbf{x}_{0})e^{ik(\mathbf{x}_{0}-\mathbf{x})^{2}/(2R)}$
where $R=P'-f$ is the distance separating the focal plane from the
the image plane, $\mathbf{x}_{0}$ the x-y coordinates of the
diffracting object in the focal plane, and $\mathbf{x}$ the x-y
coordinates in the image plane. Additionally we introduce an
effective polarizability $\alpha$ of the diffracting object. The
total intensity in the image plane is thus
\begin{eqnarray}|\alpha\Psi(\mathbf{x}_{0})|^{2}+
|\Psi_{A}(\mathbf{x})+\Psi_{B}(\mathbf{x})|^{2}+
\nonumber\\
(\Psi_{A}(\mathbf{x})+\Psi_{B}(\mathbf{x}))\cdot(\alpha\Psi(\mathbf{x}_{0})+\alpha^{\ast}\Psi^{\ast}(\mathbf{x}_{0})),
\end{eqnarray}
where $\Psi_{A}(\mathbf{x})+\Psi_{B}(\mathbf{x})$ is the field
given by Eq.~6. In the ideal case of a infinite lens we see that
not only the scattering affect the value of the intensity at the
maximum of the peaks A' and B' but that in addition there is a
background signal $|\alpha\Psi(\mathbf{x}_{0})|^{2}$ everywhere in
the image plane. These kind of objections have been presented in
particular by W.~G.~Unruh (private communications and internet
discussions) but I don't think personally that this is a
fundamental problem. Indeed it can be observed that in the
configuration used by Afshar (with both holes open and with the
grid) the effect of the wires is practically zero since the field
is practically zero on the cross-section region of the wire. There
is then no scattering light in this configuration which means that
the wire don't disturb significantly the propagation. It can be
even noted that in principle one can consider objects like atoms
which are in very good approximation punctual. Using Eq.~8 with
such object shows that the field $\Psi(\mathbf{x}_0)$ at the atoms
location equal zero when both hole A and B are open. This means a
perfect and ideal compensation of the waves reaching A' (or B')
and resulting of the atoms excitation of waves coming from A and
B. Naturally the result is different with a single aperture.
However we can not compare the single and double pinhole
experiment simply by comparing the intensity and by thinking in
terms of trajectory like in classical mechanics. Indeed in
classical mechanics a particle going through A (respectively B)
does not care about the existence of B (respectively A). It is in
that sense that we have the right to consider the sum of
probability distributions in classical physics. However this is
trivially not true in quantum physics. If we thus conclude
(wrongly) that scattering in the single aperture experiment imply
which path losses in the double-aperture experiment then we don't
consider the problem of wave duality and this is necessarily a
misleading assumption. The intuitive `` classical '' reasoning is
consequently criticizable since it uses again the ambiguous notion
of ``path'' or trajectory in the context of quantum theory based
on the principle of wave
superposition. \\
Recently Ole Steurnagel [10] analyzed numerically the effect of
the grid on the waves coming from A and B and concluded like us
that a wave contribution coming from A (respectively B) reaches
necessarily B' (respectively A') since when we open both slit the
initial pattern is practically recovered. However in [10] it is
concluded that this means a loss in the which path information.
This is again the same confusion: With optics one can consider two
configurations (here two single-aperture experiments) and add the
waves to obtain a new physical configuration. However this
wave-based reasoning doesn't mean that we have lost which path
information when both holes are open and the grid is present since
the concept of path is not univocal in this case. Bohr pertinently
observed that the concept of trajectory is senseless for an
experimentalist in the double hole experiment since we can never
observe the path without destroying fringes.

It should be finally noted in this context that in the Bohm
interpretation the photon trajectories with both holes open are
not significantly disturbed and this is in agreement with the weak
reduction of $R\sim 0.1\%$. This model is non classical since it
involves the quantum potential which is context
dependent\cite{remark} and change strogly when we open or close
one of the two holes. The Bohm model use the concept of trajectory
but since these ``real'' paths are hidden and not observable this
never contradict Bohr's statement.

The real problem in Afshar's interpretation comes from the fact
that the interference pattern is not actually completely recorded.
From the data of Afshar we would only be able to build up the part
of the interference pattern corresponding to the $R\sim 0.1\%$ of
particles absorbed by the wires. This would allow us in principle
to say how look the interference pattern close to the fringes
minima. However for the rest of the photons actually detected in
$A'$ or $B'$ we can not say anything. The situation is then
completely identical to the one without wires. The problem is that
Afshar accepts the apriori validity of wave optics. He supposes
that the small value of $R\sim 0.1\%$ is sufficient to deduce the
existence of fringes without measuring them. This is a logical
inference in the context of classical optics but this is forbidden
in quantum mechanics since we need different photons to observe
both interference and images patterns. This is not a philosophical
result but an experimental consequence of the quantum mechanics.
It is interesting to remark that a reasoning similar to the one
presented by Afshar leads to identical paradox when applied to a
configuration without lens. Indeed in that case we have fringes in
the far-field and we can deduce from the existence of the
interference pattern the existence of the two coherent sources
$A,B$ since wave optics has no other way to justify fringes like
that. Would you say that we deduced \emph{experimentally} the
existence of the source A and B? for sure not and this is the same
in the Afshar experiment: We are not able to use the same
particles to build up the different statistics in the two
complementary planes. We can even make a parallel with the
experiment of Afshar by suggesting an other simple experiment.
Suppose that instead of having in the object plane of the lens two
peaks well defined we have now as periodical sinusoidal field.
Thus from Fourier optics we will deduce that in the back focal
plane of the lens we have now two well localized peaks and that in
the image plane we have (magnified) interference fringes. We have
now completely the right to add in the back focal plane a screen
with two apertures centered in the peaks position. If the holes
are not too small this will not affect the propagation since the
field is null on the wall of the screen. Now if we observe photons
in the image plane we see fringes after statistical accumulation.
However we can conclude that the probability of presence is equal
to zero on all the screen in the focal plane since no photon are
absorbed reflected and diffracted. Do we measure in that case the
complete pattern in both complementary planes (back focal plane
and image plane)? Again the answer is not since we don't measure
what happens in the region of the apertures. It is only the
complete measurement of the two complementary patterns with the
same particles which could invoked to break the principle of
complementarity. But this is impossible as already demonstrated by
Bohr.

To conclude, in spite of Afshar's claim we still need two
experiments in order to exploit the totality of the phenomenon. As
pointed out originally by Bohr, we can not use information
associated with \emph{a same} photon event to rebuild in a
statistical way (i.e. by \emph{an accumulation of such events})
the two complementary distributions of photons in the image plane
and in the interference plane. The hypothesis of Afshar that we
only need some partial information concerning the interference
pattern in order to reconstruct the complete interference is only
based on the idea that the fringes already exist. The whole
reasoning is circular and for this reason misleading.


\begin{references}
\bibitem{Afshar}
a)S.~Afshar submitted to Physical Review Letters.\\
b)S.~Afshar, Proc.~SPIE 5866, 229-244 (July 2005).\\
c)S.~Afshar, preprint at http://www.irims.org/quant-ph/030503/.
\bibitem{Bohr1}
N.~Bohr in \emph{Albert Einstein philosopher-scientist}, edited by
P.~A.~Schilpp, (The library of living philosophers, Evanston,
1949), pp.~200-241.
\bibitem{Heisenberg}
a) W.~Heisenberg, Z.~Phys.~\textbf{43}, 172 (1927).\\
b) W.~Heisenberg, \emph{The physical principles of the quantum
theory} (University of Chicago Press,
Chicago, 1930).\\
c) W.~Heisenberg, \emph{Physics and philosophy} (Harper and
Brothers, New York, 1958).
\bibitem{Feynman}
 R.~P.~Feynman, R.~Leighton and M.~Sand, \emph{The Feynman Lectures on
 Physics} Vol.~3 (Adisson Wesley, Reading, 1965).
\bibitem{Wheeler}
a) J. A. Wheeler, \emph{in Some Strangeness in the Proportion }(H.
Woolf,
ed., AddisonWesley, 1980).\\
b) J.~A.~Wheeler, \emph{Law without law. In Quantum Theory and
Measurement} (ed. J.A. Wheeler and W.H. Zurek), Princeton
University Press pp. 182-213 (1983).
\bibitem{Englert}
a) B.~-G.~Englert, Phys.~Rev.~Lett.~\textbf{77}, 2154 (1996).\\
b) D.~M~Greenberger and A.~Yasin, Phys.~Lett.~A \textbf{128}, 391
(1988).\\c) W.~K.~Wooters, and W.~H.~Zurek,
Phys.~Rev.~\textbf{D19}, 473 (1979).\\d) G.~Jaeger, A.~Shimony,
and L.~Vaidman, Phys.~Rev.~\textbf{A51}, 54 (1995).
\bibitem{Drezet}
Bohm didn't presented a theory for photon as particle. This is
most the work of de Broglie (see for example A.~Drezet,
Opt.~Commun.~\textbf{250}, 370 (2005)).
\bibitem{Scully}
M.~O.~Scully, B.~-G.~Englert, and H.~Walther, Nature (London)
\textbf{351}, 111 (1991).
\bibitem{Scully2}
B.~-G.~Englert, M.~O.~Scully, G.~S\"{u}ssmann, and H.~Walther,
Z.~Naturforsch.~\textbf{47a}, 1175 (1992).
\bibitem{Ole}
O.~Steuernagel, arXiv: quant-ph/0512123.
\bibitem{remark}
One should observe that in Bohm's model a photon coming from $A$
goes always in $B'$ and not in $A'$ \cite{Scully2}. This is
independent of the presence of the wires (in the limit of thin
wires).

\end{references}
\end{document}